\documentclass[twocolumn,aps,pra,groupedaddress,showpacs]{revtex4}
\usepackage{epsfig,amssymb}
\def\comment#1{}\def\labell#1{\label{#1}}
\begin{document}
\title{Capacity of nonlinear bosonic systems} 
\author{Vittorio Giovannetti$^1$, Seth Lloyd$^{1,2}$, and Lorenzo
  Maccone$^1$}\affiliation{$^1$Massachusetts Institute of Technology
  -- Research Laboratory of Electronics\\$^2$Massachusetts Institute
  of Technology -- Department of Mechanical Engineering\\ 77
  Massachusetts Ave., Cambridge, MA 02139, USA}

\begin{abstract}
  We analyze the role of nonlinear Hamiltonians in bosonic channels.
  We show that the information capacity as a function of the channel
  energy is increased with respect to the corresponding linear case,
  although only when the energy used for driving the nonlinearity is
  not considered as part of the energetic cost and when dispersive
  effects are negligible.
\end{abstract}
\pacs{42.50.-p,03.67.-a,42.65.-k,89.70.+c} 
\maketitle 

Noninteracting massless bosonic systems have been the object of
extensive analysis {\cite{caves,yuen}}. Their maximum capacity in
transmitting information was derived for the noiseless case both in
the narrow-band regime (where only few frequency modes are employed)
and in the broad-band regime.  However, it is still an open question
whether nonlinearities in the system may increase these bounds:
linearity appears to be the most important assumption in all previous
derivations {\cite{caves}}. Up to now nonlinear effects have been used
in fiber optics communications to overcome practical limitations, such
as using solitons to beat dispersion or traveling wave amplifiers to
beat loss~{\cite{nonlin}}. The approach adopted in this paper is
fundamentally different from these and from others where
nonlinearities and squeezing are employed at the coding stage when
using linear channels {\cite{holevo}}.  We follow the cue of a recent
proposal {\cite{seth}} where interactions were exploited in increasing
the capacity of a qubit-chain communication line. In the case of
linear bosonic systems, the information storage capacity of a signal
divided by the time it takes for it to propagate through the medium
gives the transmission capacity of the channel.  In the presence of
nonlinearities, dispersion can affect the propagation of the signal
complicating the analysis, but an increase in the capacity can be
shown, at least when the dispersive effects are negligible. A complete
analysis of dispersion in nonlinear materials is impossible at this
stage, since the quantization of these systems has been solved only
perturbatively. The basic idea behind the enhancement we find is that
the modification of the system spectrum due to nonlinear Hamiltonians
may allow one to better employ the available energy in storing the
information: we will present some examples that exhibit such effect.
Excluding the down conversion channel (a model sufficiently accurate
to include the propagation issue --see Sec.~\ref{s:pdcbb}), all these
examples are highly idealized systems but are still indicative of the
possible nonlinearity-induced enhancements in the communication rates.
An important caveat is in order. In the physical implementations that
we have analyzed, there is no capacity enhancement if we include in
the energy balance also the energy required to create the nonlinear
Hamiltonians. This is a general characteristic of any system: if one
considers the possibility of employing all the available degrees of
freedom to encode information, then one cannot do better than the
bound obtained in the noninteracting case
{\cite{comlimit,bekensteinreply}}. However, the enhancement discussed
here is not to be underestimated since in most situations many degrees
of freedom are not usable to encode information, but can still be
employed to augment the capacity of other degrees of freedom.  A
typical example is when the sender is not able to modulate the signals
sufficiently fast to employ the full bandwidth supported by the
channel: an external pumping (such as the one involved in the
parametric down conversion case) may allow to increase the energy
devoted to the transmission modes.

We start by describing a general procedure to evaluate the capacity of
a system and we apply it to a linear bosonic system to reobtain some
known results (see Sec.~\ref{s:cap}). Such a procedure is instructive
since it emphasizes the role of the system spectrum in the capacity
calculation. We then analyze a collection of examples of nonlinear
bosonic systems in the narrow-band and wide-band regimes (see
Sec.~\ref{s:boson}): for each case we describe the capacity
enhancements over the corresponding linear systems. General
considerations on the energy balance in the information storage and on
the information propagation conclude the paper (see Sec.~\ref{s:gen}).

\section{Information capacity}\labell{s:cap}
The information capacity of a noiseless channel is defined as the
maximum number of bits that can be reliably sent per channel use. From
the Holevo bound (proven in Ref.~{\cite{yuen}} for the infinite
dimensional case), we know that it is given by the maximum of the Von
Neumann entropy $S(\varrho)=-$Tr$[\varrho\log_2\varrho]$ over all the
possible input states $\varrho$ of the channel.  In our case, since
$\varrho$ is the state of a massless bosonic field, the associated
Hilbert space is infinite dimensional and the maximum entropy is
infinite.  However, for all realistic scenarios a cut-off must be
introduced by constraining the energy required in the storage or in
the transmission, e.g.  requiring the entropy $S(\varrho)$ to be
maximized only over those states that have an average energy $E$, i.e.
\begin{eqnarray}
E=\mbox{Tr}[\varrho H]
\;\labell{ener},
\end{eqnarray}
where $H$ is the system Hamiltonian. The constrained maximization of
$S(\varrho)$ can be solved by standard variational methods (see
{\cite{bekenstein,jeff}} for examples), which entail the solution of
\begin{eqnarray}
 \delta\left\{{S(\varrho)}-\frac{\lambda}{\ln 2}{\mbox{Tr}[H\varrho]}-
\frac{\lambda'}{\ln 2}\mbox{Tr}[\varrho]\right\}=0\;\labell{variaz}
,\end{eqnarray}
where $\lambda$ and $\lambda'$ are Lagrange multipliers that take into
account the energy constraint (\ref{ener}) and the normalization
constraint Tr$[\varrho]=1$, and where the $\ln 2$ factor is introduced so
that all subsequent calculations can be performed using natural
logarithms. Eq.~(\ref{variaz}) is solved by the density 
matrix $\varrho=\exp[-\lambda H]/Z(\lambda)$ where 
\begin{eqnarray}
Z(\lambda)\equiv\mbox{Tr}[e^{-\lambda H}]\;\labell{partiz1}
\end{eqnarray}
is the partition function of the system and $\lambda$ is determined
from the constraint (\ref{ener}) by solving the equation
\begin{eqnarray}
E=-\frac{\partial}{\partial\lambda}\ln Z(\lambda)
\;\labell{lambda}.
\end{eqnarray}
The corresponding capacity is thus given by
\begin{eqnarray}
C=S\left[\exp(-\lambda H)/Z(\lambda)\right]=\left[\lambda E+\ln
  Z(\lambda)\right]/\ln 2
\;\labell{cap},
\end{eqnarray}
which means that we can evaluate the system capacity only from its
partition function $Z(\lambda)$.

In general an explicit expression for $Z(\lambda)$ is difficult to
derive, but it proves quite simple for noninteracting bosonic systems,
such as the free modes of the electromagnetic field.  In fact, in this
case the Hamiltonian is given by
\begin{eqnarray}
H=\sum_{k}\hbar\omega_{k}\;a_{k}^\dag
a_{k}\;\labell{hamilt},
\end{eqnarray}
where $\omega_k$ is the frequency of the $k$th mode and the mode
operators $a_k$ satisfy the usual commutation relations
$[a_k,a_{k'}^\dag]=\delta_{kk'}$.  Hence, the partition function is
\begin{eqnarray}
Z(\lambda)=\prod_k\sum_{n_k=0}^\infty e^{-\lambda\hbar\omega_kn_k}=
\prod_k\frac{1}{1-e^{-\lambda\hbar\omega_k}}
\;\labell{part}.
\end{eqnarray}
From Eqs.~(\ref{cap}) and (\ref{part}) it is clear that the capacity
$C$ will be ultimately determined by the spectrum $\omega_k$ of the
system. In particular, in the narrow-band case (where only a mode of
frequency $\omega$ is employed) Eq.~(\ref{cap})
gives~{\cite{caves,holevo,bekenstein}}
\begin{eqnarray}
C_{nb}=g\!\left(\frac{E}{\hbar\omega}\right),\labell{nb}
\end{eqnarray}
where $g(x)\equiv(1+x)\log_2(1+x)-x\log_2x$ for $x\neq 0$ and
$g(0)=0$.  On the other hand, in the homogeneous wide-band case (where
an infinite collection of equispaced frequencies
$\omega_k=k\delta\omega$ is employed for $k\in\mathbb{N}$)
Eq.~(\ref{cap}) gives~{\cite{yuen,caves,bekenstein,jeff}}
\begin{eqnarray}
C_{wb}\simeq\frac {\pi}{\ln 2}\sqrt{\frac{2
    E}{3\hbar\delta\omega}},\labell{wb} 
\end{eqnarray}
which is valid in the limit $\hbar\delta\omega\ll E$. When applied to
communication channels, this last equation is usually expressed in
terms of the rate $R$ (bits transmitted per unit time) and power $P$
(energy transmitted per unit time) as
\begin{eqnarray}
 R=\frac 1{\ln 2}\sqrt{\frac{\pi P}{3\hbar}}\;\labell{rate},
\end{eqnarray}
by identifying the transmission time with
$2\pi/\delta\omega$. In the next section we analyze how the capacities
$C_{nb}$ and $C_{wb}$ are modified by introducing nonlinear terms in
the system Hamiltonian.

\section{Nonlinear Hamiltonians}\labell{s:boson}
Nonlinear terms in the Hamiltonian of the electromagnetic field derive
from the interactions between the photons and the medium in which they
propagate. In this section we will employ the techniques described
above to derive the narrow-band and wide-band capacities when
quadratic nonlinearities are present. In Secs.~\ref{s:sq}, \ref{s:pdc}
and \ref{s:pdcbb} we discuss parametric down-conversion type
Hamiltonians in the narrow-band and broad-band regimes. In
Secs.~\ref{s:sw} and \ref{s:swbb} we discuss a mode swapping
interaction. All these nonlinearities arise in real-world systems from
$\chi^{(2)}$ type couplings, when one of the three fields involved in
these kinds of interactions is a strong pump field that can be
considered classical {\cite{boyd}}.  For all the cases analyzed we
present the capacity enhancement over the corresponding non
interacting Hamiltonian.

\subsection{Squeezing Hamiltonian}\labell{s:sq}
Consider the single mode described by the Hamiltonian
\begin{eqnarray}
H=\hbar\omega a^\dag a+ \hbar\xi \left[(a^\dag)^2+a^2\right]/2
+\hbar\Omega(\xi)
\;\labell{h1},
\end{eqnarray}
where $\xi$ is the squeezing parameter. We employ $|\xi|<\omega$ to
avoid Hamiltonians which are unbounded from below.  In Eq.~(\ref{h1})
the frequency $\Omega(\xi)\equiv\frac
12(\omega-\sqrt{\omega^2-\xi^2})$ has been introduced so that the
ground state of the system is null: with this choice, the average
energy $E$ is the energy associated to the mode $a$ in the nonlinear
medium. By applying the canonical transformation
\begin{eqnarray}
a=A\;\cosh\theta- A^\dag\;\sinh\theta\;,\quad\theta\equiv\frac
14\ln\left[\frac{\omega+\xi}{\omega-\xi} \right]
\;\labell{bogol},
\end{eqnarray}
the Hamiltonian $H$ is transformed to the free field form
$\hbar\sqrt{\omega^2-\xi^2}A^\dag A$, so that the derivation of the
previous section can be employed to calculate the capacity. It is thus
immediate [see Eq.~(\ref{nb})] to find the capacity of this system as
\begin{eqnarray}
C=g\left(\frac E{\hbar\sqrt{\omega^2-\xi^2}}\right)
\;\labell{c1}.
\end{eqnarray}
This quantity measures the amount of information that can be encoded
when $E$ is the total energy associated with the mode $a$, and a
nonlinear squeezing-generating term is present in the Hamiltonian:
this result is quite different from the one obtained by using squeezed
states as inputs to a linear system (see for example {\cite{holevo}}).
The capacity $C$ of Eq.~(\ref{c1}) is higher than the capacity
$C_{nb}$ of the linear case $\xi=0$ since $g(x)$ is an increasing
function (see Fig.~\ref{f:sq}).  The reason behind this enhancement is
that the nonlinearity reduces the effective frequencies of the modes
(from $\omega$ to $\sqrt{\omega^2-\xi^2}$), so that more energy levels
can now be populated with the same energy.

\begin{figure}[hbt]
\begin{center}
\epsfxsize=.8\hsize\leavevmode\epsffile{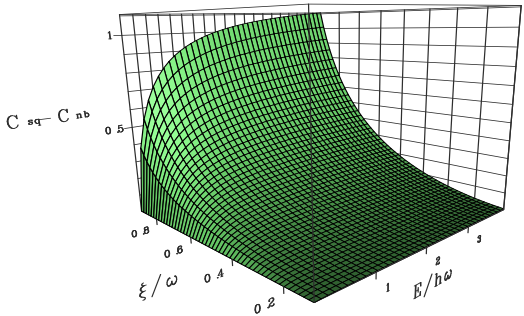}
\end{center}
\caption{Capacity increase of the squeezing Hamiltonian of
  Eq.~(\ref{h1}) as a function of the energy parameter $E/\hbar\omega$
  and of the squeezing ratio $\xi/\omega$. The increase is evident
  from the positivity of $C-C_{nb}$, where $C$ is given in
  Eq.~(\ref{c1}) and $C_{nb}$ is the free space narrow-band capacity
  of Eq.~(\ref{nb}). Notice that for high energy, the increase tends
  asymptotically to~\mbox{$-\log_2\left[1-\xi^2/\omega^2\right]/2$.}}
\labell{f:sq}\end{figure}

\subsection{Two mode parametric down conversion}\labell{s:pdc}
Consider the two interacting modes $a$ and $b$ evolved by the
Hamiltonian
\begin{eqnarray}
H=\hbar\omega(a^\dag a+b^\dag b)+\hbar\xi(a^\dag b^\dag+ab)+\hbar\Omega(\xi)
\;\labell{h2},
\end{eqnarray}
where $|\xi|<\omega$ is the coupling constant and
$\Omega(\xi)\equiv\omega-\sqrt{\omega^2-\xi^2}$ has again been
introduced to ensure that the energy ground state is null.  [This
Hamiltonian, as the previous one, possesses the structure of the
algebra of the SU(1,1) group, in the Holstein-Primakoff realization].
We can again morph the Hamiltonian to the free field form using the
two-field canonical transformation
\begin{eqnarray}
\left\{\begin{array}{l}A=a\;\cosh\theta+b^\dag\;\sinh\theta\cr
B=a^\dag\;\sinh\theta+b\;\cosh\theta\end{array}\;\right.,\quad
\theta\equiv\frac 14\ln\left[\frac{\omega+\xi}{\omega-\xi}\right]
\;\labell{c2},
\end{eqnarray}
which transforms the Hamiltonian (\ref{h2}) to the form
$\hbar\sqrt{\omega^2-\xi^2}(A^\dag A+B^\dag B)$. The capacity in this
case is 
\begin{eqnarray}
C=2\;g\left(\frac{E}{2\hbar{\sqrt{\omega^2-\xi^2}}}\right)
\;\labell{cap2},
\end{eqnarray}
which is higher than the capacity of two independent single-mode
bosonic systems, given by Eq.~(\ref{cap2}) for $\xi=0$. [The factors 2
in (\ref{cap2}) derive from the presence of the two modes $A$ and
$B$.]  The enhancement is again a consequence of the fact that the
nonlinearity reduces the effective frequency of the two modes:
$\omega\longrightarrow\sqrt{\omega^2-\xi^2}$.

\subsection{Broadband parametric down-conversion}\labell{s:pdcbb}
Here we apply the above results to the case of two wide-band modes
(signal and idler) coupled by a parametric down conversion
interaction. The interaction is mediated by a nonlinear crystal with
second order susceptibility $\chi^{(2)}$ pumped with an intense
coherent field of amplitude ${\cal E}_p$ at frequency $\omega_p$.
Assuming undepleted pumping and perfect phase matching, the
Hamiltonian at the first order in the interaction is given by
\begin{eqnarray}
H&=&\sum_k\Big[\hbar\omega_ka^\dag_ka_k+\hbar(\omega_p-\omega_k)b^\dag_kb_k
\nonumber\\
&&\qquad+\hbar\xi_k(a^\dag_kb^\dag_k+a_kb_k)+\hbar\Omega_k\Big]
\;\labell{h9},
\end{eqnarray}
where $a_k$ and $b_k$ are the mode operators of the downconverted
modes of frequency $\omega_k$ and $\omega_p-\omega_k$ respectively.
Their interaction is described by the coupling parameter
\begin{eqnarray}
\xi_k=\frac{\chi^{(2)}\pi\hbar\omega_k{\cal E}_p
}{c\epsilon_0\;n_a(\omega_k)\;n_b(\omega_p-\omega_k)}\;\Phi(\omega_k)
\;\labell{xik},
\end{eqnarray}
with $n_a$ and $n_b$ the refractive indices of the signal and idler,
and $\Phi(\omega_k)$ the phase matching function that takes into
account the spatial matching of the modes in the crystal
{\cite{boyd}}.  As usual, the frequency
$\Omega_k=[\omega_p-\sqrt{\omega_p^2-4\xi_k^2}]/2$ has been introduced
in the Hamiltonian to appropriately rescale the ground state energy.
Notice that, in contrast to the case described in Sec.~\ref{s:pdc},
the Hamiltonian (\ref{h9}) couples nondegenerate modes whose
frequencies sum up to the pump frequency $\omega_p$.  Canonical
transformations analogous to (\ref{c2}) allow us to rewrite the
Hamiltonian in the free field form
\begin{eqnarray}
H=\sum_k\left[\hbar(\omega_k-\Omega_k)A^\dag_kA_k+
\hbar(\omega_p-\omega_k-\Omega_k)B^\dag_kB_k
\right]
\;\labell{hc}.
\end{eqnarray}
With this Hamiltonian, the partition function is given by
\begin{eqnarray}
\ln
Z(\lambda)&=&\sum_k\ln\left[\frac{1}{1-e^{-\lambda\hbar(\omega_k-\Omega_k)}}\right]
\nonumber\\&+&\sum_k
\ln\left[\frac{1}{1-e^{-\lambda\hbar(\omega_p-\omega_k-\Omega_k)}}\right]
\;\labell{partiz},
\end{eqnarray}
where the two contributions are due to the signal and idler modes
respectively and the sum over $k$ is performed on all the frequencies
up to $\omega_p$. For ease of calculation, we will assume that the
coupling $\xi_k$ is acting only over a frequency band $\zeta\omega_p$
($\zeta<1$) centered around $\omega_p/2$, where it assumes the
constant value $\xi$. This choice gives a rough approximation of the
crystal phase matching function $\Phi(\omega_k)$ that prevents the
coupling of signal and idler photons when their frequencies are too
mismatched {\cite{boyd}}. In the high energy regime $E\gg
\hbar\delta\omega$, the sums in Eq.~(\ref{partiz}) can be replaced by 
frequency integrals, obtaining
\begin{eqnarray}
\ln Z(\lambda)&=&\frac{2}{\delta\omega}
\int_0^{\omega_p(1-\zeta)/2}
d\omega\;\ln\left[\frac 1{1-e^{-\lambda\hbar\omega}}\right]\nonumber\\
&+&\frac{2}{\delta\omega}
\int_{\omega_p(1-\zeta)/2}^{\omega_p(1+\zeta)/2}
d\omega\;\ln\left[\frac 1{1-e^{-\lambda\hbar(\omega-\Omega)}}\right]\nonumber\\
&+&\frac{2}{\delta\omega}
\int_{\omega_p(1+\zeta)/2}^{\omega_p}
d\omega\;\ln\left[\frac 1{1-e^{-\lambda\hbar\omega}}\right]
\;\labell{biancaneve},
\end{eqnarray}
where $\Omega\equiv[\omega_p-\sqrt{\omega_p^2-4\xi^2}]/2$.  The integrals
in Eq.~(\ref{biancaneve}) have no simple analytical solution, but we
can give a perturbative expansion in the low-interaction regime i.e.
$\epsilon\equiv 4\xi^2/\omega_p^2\ll 1$. In this limit, the result is
derived in App.~\ref{s:app} and is given by
\begin{eqnarray}
C=\frac{2\omega_p}{\delta\omega}
\left[\;c_0\!\left(\frac{E\delta\omega}{\hbar\omega_p^2}\right)
+\epsilon\;c_1\!\left(\frac{E\delta\omega}{\hbar\omega_p^2},\zeta\right)+{\cal
  O}(\epsilon^2)\right]
\;\labell{brontolo},
\end{eqnarray}
where the functions $c_0$ and $c_1$ are plotted in Fig.~\ref{f:wb}.
The zeroth order term $c_0$ in (\ref{brontolo}) gives the capacity of
two broad-band non-interacting modes with cutoff frequency $\omega_p$:
in the limit of infinite bandwidth ($\omega_p\to\infty$), this
function reaches the asymptotic behavior $C\to
C_{asym}\equiv\frac{2\pi}{\ln 2}\sqrt{E/(3\hbar\delta\omega)}$ that
corresponds to $C_{wb}$ of Eq.~(\ref{wb}) for two noninteracting
wide-band bosonic systems. To first order in $\epsilon$, the increase
in capacity due to the interaction $\xi$ is given by the value of
$c_1$, which is a positive quantity (see App.~\ref{s:app}).  It can be
shown that $c_1\to 0$ in the limit $\omega_p\to\infty$, so that in the
infinite bandwidth regime no improvement is obtained from the
interaction. In fact, an infinite continuous spectrum is invariant
under the transformation
$\omega_k\longrightarrow\sqrt{\omega_k^2-\xi_k^2}$.  As in the
non-interacting broad-band case of Eqs.~(\ref{wb}) and (\ref{rate}),
the transmission time $\tau$ of the signal can be estimated as
$2\pi/\delta\omega$. The validity of this assumption (which implies
negligible dispersion) rests on the fact that the Hamiltonian
(\ref{h9}) is valid to first order in the interaction term $\xi_k$ and
the small dispersion which derives from such term does not play any
role in the capacity {\cite{pendry,bekenstein}}.

In this calculation $E$ is the energy devoted to the signal and idler
modes: it does not include the energy spent to pump the crystal.  This
last quantity is proportional to $|{\cal E}_p|^2$ (apart from
corrections of order $\epsilon$ which derive from the coupling). In
the limit of undepleted pumping we have considered, this is a very
large contribution which overshadows $E$ and that is not directly used
to encode information.  In this respect, the system is very
inefficient in using the total available energy. However, this is not
the correct attitude to evaluate such a system: our process is
analogous to the amplification of signals where, if one considers the
total energy employed in the process (amplification plus input
energy), no gain is obtained. The correct attitude is, of course, to
consider the signals alone.

\begin{figure}[hbt]
\begin{center}
\epsfxsize=.9\hsize\leavevmode\epsffile{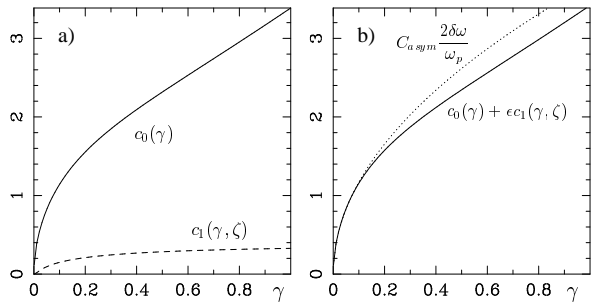}
\end{center}
\caption{a) Capacity function $c_0(\gamma)$ (continuous line) and its
  correction $c_1(\gamma,\zeta)$ (dashed line) of Eq.~(\ref{brontolo})
  with fractional coupling bandwidth $\zeta=.5$ and $\gamma\equiv
  E\delta\omega/(2\hbar\omega_p^2)$. The capacity increase
  accomplished by the nonlinearity is evident from the positivity
  of the term $c_1$. b) Comparison between the capacity $C$ of
  Eq.~(\ref{brontolo}) with $\epsilon\equiv 4\xi^2/\omega_p^2=0.1$
  (continuous line) and the asymptotic two mode rate
  $C_{asym}\equiv\frac {2\pi}{\ln
    2}\frac{\omega_p}{\delta\omega}\sqrt{2\gamma/3}$, obtained using
  an infinite frequency band (dotted line). }
\labell{f:wb}\end{figure}

\subsection{Swapping Hamiltonian}\labell{s:sw}
Consider $N$ modes that are pairwise coupled through the
Hamiltonian
\begin{eqnarray}
H&=&\sum_{j=1}^N\hbar\omega a^\dag_ja_j+\sum_{j\neq j'}
\Lambda_{jj'}a^\dag_j a_{j'}\nonumber\\&=&
\hbar\vec a\:^\dag\cdot(\omega\openone+{\Lambda})\cdot\vec a
\;\labell{h5},
\end{eqnarray}
where $\vec a$ is the column vector containing the annihilation
operators $a_j$ of the $N$ modes and $\Lambda$ is an $N\times N$
symmetric real matrix with null diagonal. This Hamiltonian describes
$N$ modes $a_j$ whose photons have a probability amplitude
$\Lambda_{jj'}$ to be swapped into the mode $a_{j'}$. By performing a
canonical transformation on all the mode operators, it is possible to
rewrite Eq.~(\ref{h5})  in the free field form
\begin{eqnarray}
H=\sum_{j=1}^N\hbar(\omega+\lambda_j)A_j^\dag A_j
\;\labell{hh},
\end{eqnarray}
where the $\lambda_j$s are the $N$ eigenvalues of $\Lambda$. Two
conditions must be satisfied: the positivity of $H$ requires that
$\lambda_j\leqslant\omega$ for all $j$, and, since the diagonalization
of $\Lambda$ must preserve its trace, $\sum_j\lambda_j=0$. The
capacity of this system can now be easily computed as
{\cite{yuen,caves,bekenstein}}
\begin{eqnarray}
C=\max_{e_1,e_2,\cdots,e_N}\left[\sum_{j=1}^N
g\left(\frac{e_j}{\hbar(\omega+\lambda_j)}\right)
\right]
\;\labell{csw},
\end{eqnarray}
where the maximum must be evaluated under the requirement that
$\sum_je_j=E$.  In the simple case of $N=2$ (where
$\lambda_1=-\lambda_2\equiv\xi$), the maximization (\ref{csw}) can be
easily performed numerically: the increase in capacity over the two
mode non interacting case is presented in Fig.~\ref{f:sw}.  In this
case, in the strong coupling regime ($|\xi|\to\omega$) the capacity
diverges as $\log_2[E/(\hbar(\omega-|\xi|))]$: this corresponds to
employing for the information storage only the lowest frequency mode
among $A_1$ and $A_2$.

\begin{figure}[hb]
\begin{center}
\epsfxsize=.8\hsize\leavevmode\epsffile{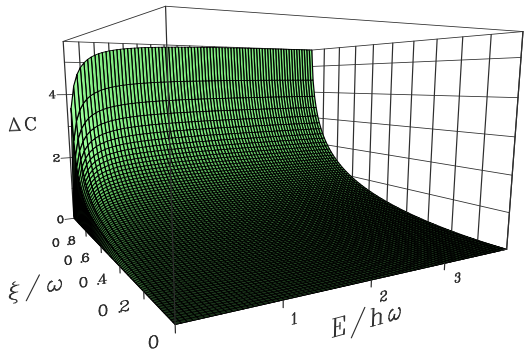}
\end{center}
\caption{Capacity increase of the swapping Hamiltonian as a function
  of the energy parameter $E/\hbar\omega$ and of the coupling ratio
  $\xi/\omega$, given by plotting $\Delta C\equiv
  C-2g(E/(2\hbar\omega))$ (the first term being the capacity $C$ of
  Eq.~(\ref{csw}) and the second the capacity of two non interacting
  modes that share an energy $E$). } \labell{f:sw}\end{figure}

\subsection{Broadband swapping Hamiltonian}\labell{s:swbb}
A generalization of the Hamiltonian (\ref{h5}) given in the previous
subsection can be obtained by considering $N$ parallel broadband modes
in which the coupling joins all the modes with the same frequency,
namely,
\begin{eqnarray}
H&=&\sum_k\sum_{j=1}^N\hbar\omega_k a^\dag_{jk}
a_{jk}
\nonumber\\&+&\sum_k\sum_{j\neq j'}
\Lambda^{(k)}_{jj'}a^\dag_{jk} a_{j'k}
\;\labell{h6},
\end{eqnarray}
where $a_{jk}$ is the annihilation operator of the $j$th system with
frequency $\omega_k$.  This Hamiltonian can be taken to a free field
form using the procedure of the previous subsection, so that
\begin{eqnarray}
H=\sum_k\sum_{j=1}^N\hbar(\omega_k+\lambda_{jk})A_{jk}^\dag
A_{jk}
\;\labell{mi},
\end{eqnarray}
where $\lambda_{jk}$ is the $j$th eigenvalue of the matrix
$\Lambda^{(k)}$. Expressed in terms of the normal modes $A_{jk}$, the
Hamiltonian (\ref{mi}) describes $N$ independent wideband modes.
Since in this case the maximization of the form (\ref{csw}) is
complicated, we show an increase in capacity by considering the
following choice of the coupling constants (which may well not be the
optimal one):
\begin{eqnarray}
\lambda_{jk}&=&-\omega_k\;{r}\qquad\mbox{ for
}j=1,\cdots, N-1\;\labell{fa}\\
\lambda_{Nk}&=&\omega_k(N-1)\;{r}
\;\labell{sol},
\end{eqnarray}
where $0\leqslant {r}\leqslant 1$.  The choices (\ref{fa}) and
(\ref{sol}) automatically guarantee both the positivity condition on
the Hamiltonian and the trace preservation condition on the
$\Lambda^{(k)}$ matrices. With this choice, the Hamiltonian (\ref{mi})
becomes
\begin{eqnarray}
H&=&{(1-{r})}\sum_k\sum_{j=1}^{N-1}
\hbar\omega_kA_{jk}^\dag
A_{jk}\nonumber\\
&&+[1+(N-1)\;{{r}}]\sum_k\hbar\omega_kA_{Nk}^\dag
A_{Nk}
\;\labell{la}.
\end{eqnarray}
Essentially we have chosen a coupling that contracts by a factor
$1-{r}$ the frequencies of the first $N-1$ normal-modes and stretches
by a factor $[1+(N-1)\;{{r}}]$ the frequency of the $N$th one.
Choosing ${r}\to 1$, we can increase the capacity {\it ad libitum}
over the case of $N$ non-interacting systems. In fact, in this limit a
straightforward application of the wideband calculation of
Eq.~(\ref{wb}), gives ${C}\simeq\frac{N-1}{1-{r}}{C_{wb}}$.
%
This result must be compared with the case in which the $N$ wideband
modes are independent where the capacity scales as $\sqrt{N}C_{wb}$.
Clearly an arbitrary increase in capacity is gained as ${r}\to 1$.

\section{General considerations}\labell{s:gen}
In the previous sections we have derived the capacity of various types
of nonlinear bosonic systems. The common feature of all these results
is that the nonlinearities were used to reshape the spectrum by
compressing it to lower frequencies, where it is energetically cheaper
to encode information. A couple of remarks are in order. First of all,
in our calculations the mean energy $E$ represents the energy of the
modes {\em in} the system, whereas customarily one considers the input
energy. Our choice is motivated by the fact that nonlinear systems
dispense energy to the information bearing modes, so that the input
energy does not necessarily coincide with the amount of energy that
the medium where information is encoded needs to sustain. This last
quantity is a practically relevant one for the cases in which the
degrees of freedom used to encode information cannot handle high
energies (as for example optical fibers {\cite{naturenonlin}}).
Notice that, in some of the cases we have studied (the example of
Sec.~\ref{s:pdcbb} in particular), the nonlinearity is achieved by
supplying the system with an external energy source in the form of an
intense coherent beam. If one were to take into account also this
contribution in the energy balance, then no capacity enhancement would
be evident, since the pumping energy is used only indirectly to store
the information. However, it is not unwarranted to exclude the pump
from the energy balance, since it is only used to set up the required
Hamiltonian and not directly employed in the information processing.
Finally, our analysis is limited to the noiseless case. In the
presence of noise in place of the von Neumann entropy in
Eq.~(\ref{variaz}), one would have to consider the Holevo
information~{\cite{hsw}}.  This is a highly demanding problem because
of the yet unknown additivity properties of this quantity. At least in
the case of linear bosonic systems, the capacity in the presence of
noise was studied in {\cite{holevo,nostro}}.

This work was funded by the ARDA, NRO, NSF, and by ARO under a MURI
program.

\appendix
\section{}\labell{s:app}
In this appendix we derive formula (\ref{brontolo}) for the broadband
parametric down-conversion Hamiltonian. 
In the low-interaction regime $\epsilon\ll 1$, Eq.~(\ref{biancaneve})
becomes
\begin{eqnarray}
\ln Z(\lambda)=\frac{2\omega_p}{\delta\omega}\left[
f_0(\beta)+\epsilon f_1(\beta,\zeta)+{\cal
  O}(\epsilon^2)
\right]
\;\labell{giuggiolo},
\end{eqnarray}
where $\beta\equiv\lambda\hbar\omega_p$ and 
\begin{eqnarray}
f_0(\beta)&\equiv&\frac 1\beta\int_0^\beta dx\;\ln\left[\frac 1{1-e^{-x}}\right]
\;\labell{f0}\\
f_1(\beta,\zeta)&\equiv&\frac 14\ln\left[\frac{1-e^{-\beta(1+\zeta)/2}}
{1-e^{-\beta(1-\zeta)/2}}\right]\labell{f1}\;.
\end{eqnarray}
Notice that the zeroth order term $f_0$ in the expansion
(\ref{giuggiolo}) corresponds to the partition function of two
broadband modes (signal and idler) with cut-off frequency $\omega_p$.
Replacing Eq.~(\ref{giuggiolo}) into the energy constraint
(\ref{lambda}) we can find the value of the Lagrange multiplier
$\lambda$, contained in the parameter $\beta$, by solving the equation
\begin{eqnarray}
\frac{\partial f_0(\beta)}{\partial\beta}+\epsilon
\frac{\partial f_1(\beta,\zeta) }{\partial\beta}
=-\gamma
\;\labell{principe},
\end{eqnarray}
where $\gamma\equiv E\delta\omega/(2\hbar\omega_p^2)$ is a
dimensionless quantity.  By expanding the solution $\beta$ for small
$\epsilon$ as $\beta=\beta_0+\epsilon\beta_1$, it follows that
\begin{eqnarray}
\frac{\partial f_0(\beta_0)}{\partial\beta}=-\gamma\;,\qquad\labell{do}\\
\beta_1=-\frac{\partial
  f_1(\beta_0,\zeta)}{\partial\beta}\Bigg/\frac{\partial^2
  f_0(\beta_0)}{\partial\beta^2} \;\labell{re}.
\end{eqnarray}
These two equations can be numerically solved for any value of
$\gamma$. Replacing the solution in (\ref{giuggiolo}) and using the
capacity formula (\ref{cap}) we can evaluate the parametric
down-conversion capacity as reported in Eq.~(\ref{brontolo}) where
\begin{eqnarray}
c_0(\gamma)\equiv\left[\beta_0\gamma+f_0(\beta_0)\right]/\ln 2
\;\labell{mi1},\\
c_1(\gamma,\zeta)\equiv f_1(\beta_0,\zeta)/\ln 2\;.\quad
\end{eqnarray}
Both these functions depend on the system energy only through the
quantity $\gamma$.

\end{document}